# In-situ integrated nonlocal Huygens metasurfaces enable robust vortex beams from multimode fibers

Mingke Jin[1†], Jieyu Zheng[1†], Chaoyang Wang[1[+]], Yuhao Fang[1], Gongkai Zhang[2], Min Qiu[1], Jingyi Tian[1*]

[1]Zhejiang Key Laboratory of 3D Micro/Nano Fabrication and Characterization, Department of Electronic Science and Engineering, School of Engineering, Westlake University, Hangzhou, Zhejiang, China 310024.
[2]Westlake Center for Micro/Nano Fabrication, Westlake University, Hangzhou, Zhejiang, China 310024.
[†]These authors contributed equally to this work.
[+] Present address: Eastern Institute of Technology, Ningbo, Zhejiang, China 315200.
*Corresponding author. E-mail: tianjingyi@westlake.edu.cn

## Abstract
A fundamental hurdle for multimode fiber (MMF) integrated photonics is achieving robust and deterministic wavefront shaping against fiber perturbation, which induces random modal interference and scrambles the real-space output into speckle patterns. Existing wavefront shaping methods typically rely on pixelated phase/polarization pre-compensation, rendering them highly sensitive to fiber deformation. Here, we show that in-situ integration of a silicon nonlocal metasurface onto an MMF end facet enables direct conversion of arbitrary speckle patterns into optical vortex beams without pre-compensation. The metasurface supports overlapped Mie resonances and nonlocal bound states in the continuum (BICs), yielding high-transmission bands by satisfying the generalized nonlocal Huygens condition. Through momentum-space coupling between MMF modes and Huygens-BICs with topological polarization-vortex nature, the output maintains doughnut-shaped profiles under static and dynamic perturbations, demonstrating exceptional robustness against fiber deformation. This approach provides a compact, alignment-tolerant platform for robust structured light generation in MMFs for endoscopy, optical trapping and communications.

## Main

Optical fibers[1,2], as the cornerstone of optical information and energy transmission, offer low loss, strong light confinement, long interaction lengths and high flexibility, enabling wide applications in optical communications[3,4], endoscopic imaging[5,6], fiber sensing[7,8], fiber lasers[9,10], and lab-on-fiber technology[11,12,13]. The integration of metasurfaces onto fiber end facets has recently given rise to 'metafibers'[14,15], which extend conventional fiber functionality by allowing precise wavefront control at the distal tip. To date, however, metafiber research has almost exclusively focused on single-mode fibers (SMFs)[16-23], where the well-defined Gaussian output can be readily manipulated by real-space phase or polarization modulation. The MMFs, which support hundreds to thousands of spatial modes as potential independent channels[24], remains largely unexplored. This is a significant opportunity, as MMFs are pivotal for real-time high-resolution micro-endoscopy[25,26], space-division multiplexing[27], high-performance optical communication[28], all-optical image transportation[29] and high-dimensional optical manipulation[30,31]. However, the feature that enables this high capacity, namely the large core diameter of MMFs, also introduces a fundamental challenge for precise and robust light manipulation[32]. Specifically, the large core diameter of MMFs allows light to propagate in numerous modes with distinct propagation constants[33] and transverse field distributions. Random perturbations, such as bending, non-uniform stress, or core-diameter fluctuations, inevitably alter mode coupling, producing a random and unstable speckle-like output[34,35]. Consequently, conventional real-space pixelated phase- or polarization-type diffractive optical elements[36,37] implemented on single-mode (SM) fiber end facets cannot be directly applied to MMFs, as their performance require precise pixelated pre-compensation and degrade severely under environmental variations, bending, or twisting[38,39]. This poses a fundamental challenge for MMF-based integrated photonic devices: achieving high-performance, robust light manipulation under realistic operating conditions.

In this context, nonlocal metasurfaces[40-44] emerge as a compelling solution. Nonlocal metasurfaces operate through collective resonant modes sustained across the entire periodic lattice, with their optical signatures intrinsically encoded in the momentum space and protected by global symmetries, conferring exceptional robustness against real-space local perturbations. Therefore, in-situ integration of a nonlocal metasurface onto the end facet of an MMF offers a promising route to alleviate the robustness issue inherent to MMF-integrated devices, enabling reliable signal transmission and optical field manipulation. Realizing such in-situ nonlocal metasurface integration, however, faces substantial fabrication challenges. The narrow and flexible geometry of optical fibers is inherently incompatible with standard CMOS fabrication, posing significant difficulties in device fixation, alignment and on-fiber integration. Common techniques for fabricating functional metafibers, such as 3D nano-printing[45,46], focused ion beam (FIB) etching[20,23], and split-transfer processes[21,47], have intrinsic limitations. 3D nano-printing employs low-refractive-index materials, which cannot support the strong electromagnetic resonances and extreme light confinement. FIB etching suffers from ion implanting and increased recasting with depth, leading to high loss and uncontrolled morphology, which complicates the fabrication of high-aspect-ratio nanostructures on fiber end facets. The split-transfer process faces inherent alignment inaccuracies, impeding the miniaturization and integration of metafiber photonic devices. Therefore, directly integrating high-refractive-index, low-loss dielectric metasurfaces onto MMF end facets as a compact transmissive photonic device remains a critical challenge.

To address both the robustness and integration challenges, we propose a C-band optical vortex generator based on a silicon (Si) nonlocal metasurface fabricated directly onto an MMF end facet. The dielectric metasurface supports overlapped broadband Mie resonances and multiple high-quality-factor BICs, yielding high transmission bands by satisfying the generalized nonlocal Huygens condition for fiber-based applications. Unlike SM metafibers that rely on real-space pixelated phase modulation for vortex generation, our MMF-integrated design employs a real-space uniform nanostructure, effectively bypassing the drawback of high-sensitivity to local perturbations, such as local defects on the fiber end facet, mode distortions, or positional misalignments in pixelated approaches. This scheme operates via momentum-space coupling between fiber modes and nonlocal Huygens-BICs that possess a topological polarization-vortex nature. As a consequence, this directly converts speckle patterns into a momentum-space polarization vortex beam (or equivalently a pair of chiral vortices with opposite topological charges) without any pre-compensation. Using electron beam lithography (EBL) and inductively coupled plasma (ICP) etching with CMOS-compatible holders and baseplates for high-precision fiber alignment, we successfully fabricate a subwavelength Si metasurface onto the end facet of a commercial MMF. The generation of vortex beams is experimentally confirmed by far-field unpolarized and polarized intensity map around the BIC wavelength in momentum space. Moreover, the donut-shape radiation pattern remains nearly unchanged under external fiber bending and twisting perturbations, confirming the robustness of the metafiber. Our results demonstrate a compact and versatile platform for MMF-based light manipulation, holding high promise for applications in optical information processing, structured light generation, miniature endoscopy, optical trapping, sensing and signal transmission.

## Principle and design

Fig. 1a presents the concept of the proposed MMFs integrated with nonlocal

Huygens metasurfaces for realizing the optical vortex beam generation at the communication band. The system employs a commercial MMF with a core diameter of 100 μm and a numerical aperture (NA) of 0.22. Unpolarized light is incident into the MMF, and its end facet outputs an arbitrary speckle pattern owing to the distinct propagation constants and distributions of the various guided modes in the MMF. This disordered field pattern serves as the excitation for the nonlocal metasurface, which comprises two-dimensional periodic high-index nanostructures. Acting as a momentum-space wavefront-shaping device, the nonlocal transmissive metasurface maps the MMF outputs onto its momentum-space states of polarization (SoPs), which manifest in the far field as vortex beams.

Notably, conventional SM metafibers typically generate vortex beams by imprinting a helical phase profile directly onto the outgoing Gaussian beam in real space (Fig. 1b). In MMFs, however, the speckled output with arbitrary phase and polarization distributions precludes the use of standard real-space pixelated phase- or polarization-type diffractive elements. Our MMF-integrated metadevice instead harnesses this random speckle field as the input and, via coupling with nonlocal metasurface modes, reconstructs it into momentum-space vortex beams (Fig. 1c). This approach leverages momentum-space degrees of freedom, enabling efficient, alignment-tolerant and robust wavefront reconstruction with MMFs.

Our metasurface design is guided by two essential requirements: (1) topological robustness against real-space perturbations, (2) transmission-mode operation for direct fiber integration. To meet these demands, we employ a nonlocal metasurface consisting of a square lattice of triangular Si meta-atoms with broken in-plane inversion symmetry. This geometry is deliberately chosen because symmetry breaking converts ideal BICs, which are dark at $\Gamma$ and inaccessible from the far field, into radiative $q$-BICs with controlled linewidths, while simultaneously creating momentum-space polarization vortices carrying topological charges. These topological resonances, when coupled to the random speckle field of an MMF, provide a direct pathway from a disordered real-space input to a well-defined momentum-space output.

The nonlocal metasurface supports multiple $q$-BICs and Mie resonances in the spectral region of interest. Fig. 2 presents two structural variants (meta-1 and meta-2, with meta-atom heights of 259 nm and 518 nm, respectively), illustrating how the meta-atom height controls the spectral overlap between the $q$-BICs supported by the lattice and the broad Mie resonances supported by the meta-atoms, which is a tuning knob for achieving high-transmission by satisfying the generalized Huygens condition[48]. Both geometries support two $q$-BICs near the C-band (Figs. 2a and 2b). Near-field analysis identifies $q$-BIC 1 as a magnetic-dipole mode (in-plane current loop, Fig. 2c) and $q$-BIC 2 as an electric-dipole mode (out-of-plane oscillation, Fig. 2e). Their far-field polarization distributions in momentum space form vortex patterns[41] with topological charges of +1 and −1, as expected (Figs. 2d and 2f).

The high-transmission behavior emerges from interference between the narrow vertical dipole $q$-BICs and the broad Mie resonances in the far field. The angle-resolved transmission spectra and multipole decompositions of meta-1 (Figs. 2g and 2i) and meta-2 (Figs. 2h and 2j) are calculated under p and s-polarization excitation (Details of the numerical simulations are provided in the Methods and Supplementary Note 1). Notably, as the meta-atom height increases, the narrow transmission dip associated with the $q$-BIC 1 is converted into a transmission peak (Fig. 2g → Fig. 2h), due to coupling between the narrow magnetic-dipole $q$-BIC (out-of-plane magnetic dipole mode) supported by the lattice and broad Mie resonances (in-plane electric and magnetic dipole modes) supported by the Si meta-atoms. In contrast, the high-transmission peak associated with $q$-BIC 2 persists, resulting from the sustained spectral overlap and interference between the electric-dipole $q$-BIC and the Mie resonances (Figs. 2i and 2j) as the height varies.

We first validate the design on a planar quartz substrate (100 μm × 100 μm footprint). The top-view and side-view scanning electron microscopy (SEM) images of the metasurface sample are shown in Figs. 3a and 3b. A commercial angle-resolved micro-spectrometer (ARMS, Ideaoptics, China) is used to obtain full-angle information in a single image by mapping different emission angles from the sample to different positions on the objective's back focal plane. A schematic diagram of the detailed measurement optical path is shown in Fig. 3c. Figs. 3d and 3e shows the measured angle-resolved reflection and transmission spectra within the angular range of ±10°. The measured high-transmission bands of $q$-BIC 1 and $q$-BIC 2 are clearly resolved and in excellent agreement with the simulation (Figs. S3). The far-field doughnut profile of $q$-BIC 1 within 12° emission angle (Fig. 3c, inset) confirms the vortex nature of the emission. These results confirm that our metasurface, featuring multiple Huygens-BIC modes centered at different polarization vortices, provides a viable platform for high-transmission vortex generation.

**On-fiber integration and characterization**

We next integrate this nonlocal Huygens metasurface directly onto the end-facet of a commercial MMF to achieve on-fiber wavefront manipulation. The fabrication is carried out using a standard CMOS process, complemented by CMOS-compatible holders and baseplates to ensure high-precision mechanical alignment during integration (Figs. 4a–4e). To avoid thermal damage to the optical fiber, amorphous silicon (a-Si) films are deposited onto the fiber end-facet via electron beam evaporation (EBE). The measured refractive index of these films is provided in Fig. S9b. To make sure that the physical dimensions of the fiber are compatible with the EBL chamber, we use a commercial SEM (Zeiss, Gemini360) equipped with a Raith pattern generator to expose the metasurface pattern. A chromium (Cr) hard mask, combined with ICP etching, is then employed to produce Si meta-atoms with vertical sidewalls and well-defined morphologies. Detailed fabrication procedures are provided in Methods. The proposed on-fiber integration method lays a solid technological foundation for the development of ultra-compact, low-loss, multifunctional transmissive metafiber devices.

The optical photograph of the MMF-integrated metadevice is presented in Fig. 4f, along with the top-view and side-view SEM images of the on-fiber nonlocal metasurface in Figs. 4g and 4h. The detailed experimental setup is shown in Fig. 4i. The experimental angle-resolved reflection and transmission spectra of the on-fiber nonlocal metasurface are respectively shown in Figs. 4j and 4k, where the blue and orange dashed lines illustrate the $q$-BIC 1 and $q$-BIC 2 bands. During ICP etching of triangular Si meta-atoms, the associated thermal effect leads to inward recess of the meta-atom sidewalls, resulting in a deviation from the ideal vertical sidewall profile. To account for this fabrication-induced non-ideality and accurately characterize the optical performance of the on-fiber nonlocal metasurface, we calculated the angle-resolved transmission and reflection spectra using the actual structural parameters, as shown in Figs. S11a and S11b (the actual computational model and corresponding optical band of the triangular meta-atom are shown in Supplementary Note 3). We can find that the experimentally measured angular spectra of the nonlocal metasurface on the MMF end-facet agree well with the calculated results. The measured transmission spectrum in Fig. 4j confirms that $q$-BIC 2 exhibits a clear high-transmission behavior, while it should be noted that the suppression of the transmission near $q$-BIC 1 in the experiment is attributable to the scattering loss and residual Si layer on the MMF end face.

**Momentum-space vortex beam generation**

As schematically shown in Fig. 5a, the metadevice exploits momentum-space coupling between the guided modes of the MMF and the nonlocal metasurface, which excites $k$-dependent SoPs in the vicinity of $q$-BIC 2 and ultimately produces vortices in the far field. Owing to the spin–orbit locking inherent in the BIC topological configuration[41,49], the two spin components ($\sigma_\pm$) acquire phase vortices with opposite winding numbers. For the specific $q$-BIC 2 resonance (Fig. 5b(i)), which carries a topological charge of −1, the spin–orbit locking dictates that the left circular polarization (LCP) component exhibits a vortex with $l = -1$ (Fig. 5b(ii)), while the right circular polarization (RCP) component possesses a vortex with $l = +1$ (Fig. 5b(iii)).

To directly visualize the vortex generation around $q$-BIC 2, we switch the spectrometer from spectral mode (Fig. 4i) to imaging mode. A narrowband filter centered at 1520 nm (bandwidth 10 nm, marked by white dashed lines in Fig. 4j) is used to isolate the transmitted beam, which is then Fourier-transformed into the far field by an objective lens. As presented in Fig. 5c(i), the far-field intensity profiles exhibit a clear doughnut shape, a hallmark of vortex beams. Moreover, the momentum-space topological polarization configuration of $q$-BIC 2 is probed through linear polarization imaging (Figs. 5c(ii)–(v)), which unambiguously reveals the winding of the electric-field vector around the singularity, consistent with a −1 topological charge. Finally, to verify the spin–orbit locking associated with $q$-BIC 2, we separate the LCP and RCP components and observe that both components display doughnut-shaped far-field radiation patterns (Figs. 5c(vi)–(vii)).

**Robust vortex generation against fiber deformation**

In MMFs, bending and twisting inherently induce phase distortions and polarization crosstalk, which severely compromise the performance of conventional MMF-based integrated photonic devices. By contrast, our nonlocal metasurface operates intrinsically in momentum space via mode coupling. Its topological protection against local real-space perturbations ensures that the wavefront shaping remains effective even under significant fiber deformation. To quantitatively assess this robustness, we deliberately bend the MMF body into a series of well-defined configurations (schematic in Fig. 6a) and measure the far-field intensity distributions at the bending angles of 0°, 45°, 90°, 135° and 180°. As shown in Fig. 6b, the characteristic doughnut-shaped vortex profiles generated by the on-fiber nonlocal Huygens metasurface are essentially preserved across all bending angles, demonstrating the excellent robustness of the MMF-based metadevice against fiber deformation. In stark contrast, the output of a bare (metasurface-free) MMF under the same bending conditions exhibits randomly varying speckle patterns (Fig. 6c), as expected from uncontrolled mode interference.

To further test the robustness limit, we subject the MMF to complex, randomly combined deformations—including twisting, bending and stretching—as depicted in Fig. 6d. The doughnut-shaped intensity profiles from the metasurface-integrated MMF remain remarkably stable across all perturbation states (Fig. 6e), while the bare MMF output continues to fluctuate randomly (Fig. 6f). These results demonstrate that our momentum-space approach provides a practical and reliable solution for stable, high-fidelity wavefront shaping in MMFs, even in harsh and unpredictable environments.

Beyond static deformations, real-world applications often involve continuous dynamic perturbations, including sustained bending from physiological motions such as heartbeat or respiration during endoscopic procedures[25], or complex strains (stretching, torsion, compression) encountered in soft robotic manipulations[50]. To evaluate the time-averaged stability of the generated vortex fields under such dynamic conditions, we record the far-field intensity distribution at the Fourier plane over a 1.8 s exposure period while the MMF-based metadevice is continuously and randomly bent and twisted (Fig. 6g; $S_1$, $S_2$ and $S_3$ denote three representative snapshots during the stochastic process). The integrated intensity patterns confirm that the vortex mode retains high modal fidelity in the time-averaged sense, with even better doughnut profile. This robustness, together with alignment tolerance, makes our device a practical ‘plug-and-play’ solution for MMF-based endoscopy, optical trapping and communication.

## Conclusions

In summary, we have demonstrated a robust C-band optical vortex generator by in-situ integrating a high-index dielectric nonlocal Huygens metasurface directly onto the end facet of a commercial MMF. The designed metasurface supports multiple $q$-BICs, associated with momentum-space polarization vortices. These $q$-BIC modes, when hybridized with broad Mie resonances of high-index meta-atoms and fulfilling generalized Huygens condition, give rise to high-transmission bands for fiber-based optical devices. By exploiting momentum-space coupling between the guided modes of the MMF and the nonlocal Huygens-BICs of the metasurface, our approach directly converts the arbitrary, unstable output of the MMF into well-defined vortex beams without any pre-compensation. The momentum-space operation endows the metafiber with exceptional robustness against real-space local perturbations.

Beyond vortex generation, the demonstrated concept can be readily extended to other forms of momentum-space light shaping, such as structured beam generation, beam shifting or complex wavefront engineering. The high-Q dielectric resonances also open avenues for ultrasensitive sensing, low-threshold lasers, and nonlinear frequency conversion on a fiber tip. Moreover, the integration of active materials in the future could enable dynamic tunability of the optical modes, leading to reconfigurable metafiber devices for adaptive optics and intelligent light manipulation. Overall, our work provides a robust and versatile foundation for ultra-compact, multifunctional metafiber devices, with promising implications for miniature endoscopy, optical trapping and high-dimensional optical communication.

## Methods

### In-situ fabrication of dielectric nonlocal Huygens metasurfaces on the fiber end facet

First, the outer protective jacket of a commercial MMF was stripped to expose the fiber patch cord, which was then cleaned. Subsequently, an a-Si film was deposited onto the fiber end facet by EBE process. The refractive index and thickness of the deposited film were calibrated using a reference film simultaneously deposited on an adjacent quartz substrate. After that, the fiber was fixed in a custom holder, and a 120-nm-thick layer of electron-beam resist PMMA (baked at 180 °C) followed by a conductive polymer ARPC 5090.05 (baked at 90 °C) was sequentially spin-coated onto the fiber end face. Subsequently, the fiber was mounted onto a designed holder and transferred into an EBL system, wherein the metasurface pattern was inscribed into the resist layer with the exposure dose of 550 $\mu$C/cm$^2$. After EBL exposure, the conductive polymer was removed with deionized water, followed by immersing in a developer solution (MIBK: IPA = 1: 3) for 38 seconds and then in isopropyl alcohol (IPA) for 1 minute. Subsequently, a Cr hard mask layer was deposited onto the patterned nonlocal metasurface, and then the metasurface pattern on the MMF was immersed in acetone solution for lift-off process. Next, the patterned fiber was transferred to a dedicated etching holder, and the a-Si film on the MMF-end-facet was etched using ICP process with a gas mixture of $SF_6$ and $CHF_3$. After ICP etching, the fiber was immersed in an etchant solution to remove the residual Cr layer. Finally, the fiber-integrated nonlocal metasurface was spliced to a bare fiber patch cord to achieve in-situ fabrication of dielectric metasurfaces on the fiber end facet.

**Numerical simulations**

The eigenmode simulations for the band structure, Q-factor, near-field distributions of the electric field amplitude and Stokes parameters were conducted by using COMSOL Multiphysics. In these simulations, Floquet periodic boundary conditions were imposed in the transverse direction, while perfectly matched layers were adopted in the $z$-direction. The far-field polarization states were reconstructed by calculating the Stokes parameters, and the azimuthal angle $\psi$ and polar angle $\chi$, which were obtained from the average electric field ($c_x$, $c_y$) at a plane far away from the triangular nanostructure. Detailed calculation methods are provided in Supplementary Note 1. Angle-resolved reflection and transmission spectra were simulated by using rigorous coupled wave analysis within the Ansys Lumerical software. In these calculations, the electromagnetic field was expanded via spatial Fourier series with continuity boundary conditions at interfaces, and the solved complex amplitudes of each diffraction order yielded the angle-resolved reflectance and transmittance spectra. Besides, the refractive index of the Si nanostructure used in the above simulations was obtained from ellipsometry measurements of the thin film on quartz substrate (Fig. S8).

**Data availability**

The data supporting the findings of this work are available both within this article and Supplementary Information. The raw data of this study are available from the corresponding authors upon request.

## Acknowledgements

The authors thank Westlake Center for Micro/Nano Fabrication and Instrumentation and Service Center for Molecular Sciences at Westlake University for the facility support and technical assistance. The author thanks Yanan Yang from the Materials Center of Tsinghua University for the assistance of the angle-resolved spectroscopy.

## Funding Statement

This work was supported by the National Natural Science Foundation of China (Grant No. 62405249 and Grant No. 62505252), the Research Center for Industries of the Future (RCIF) at Westlake University (Grant No. 210000006022305) and the Start-up fund of Westlake University (Grant No. 103110786022301).

## Author contributions

M.J., and J.Z. contributed equally to this work. M.J. and J.T. conceived the idea and supervised the project. M.J., J.Z. and J.T. optimized the geometrical parameters of the meta-atoms and designed the BIC metasurface sample, M.J. fabricated the BIC metasurface onto the end facet of MMF and characterized the optical properties of the MMF-based metadevice. C. W. participated in the design of the fiber holders used in the fabrication process and Y. F. contributed to the fiber splicing process. G. Z. provided facility support and technical assistance for the ICP etching process. M. J.

wrote the first version of the manuscript that was edited by the rest of the authors. All authors contributed to the data analysis.

## Competing interests

The authors declare no competing interests.

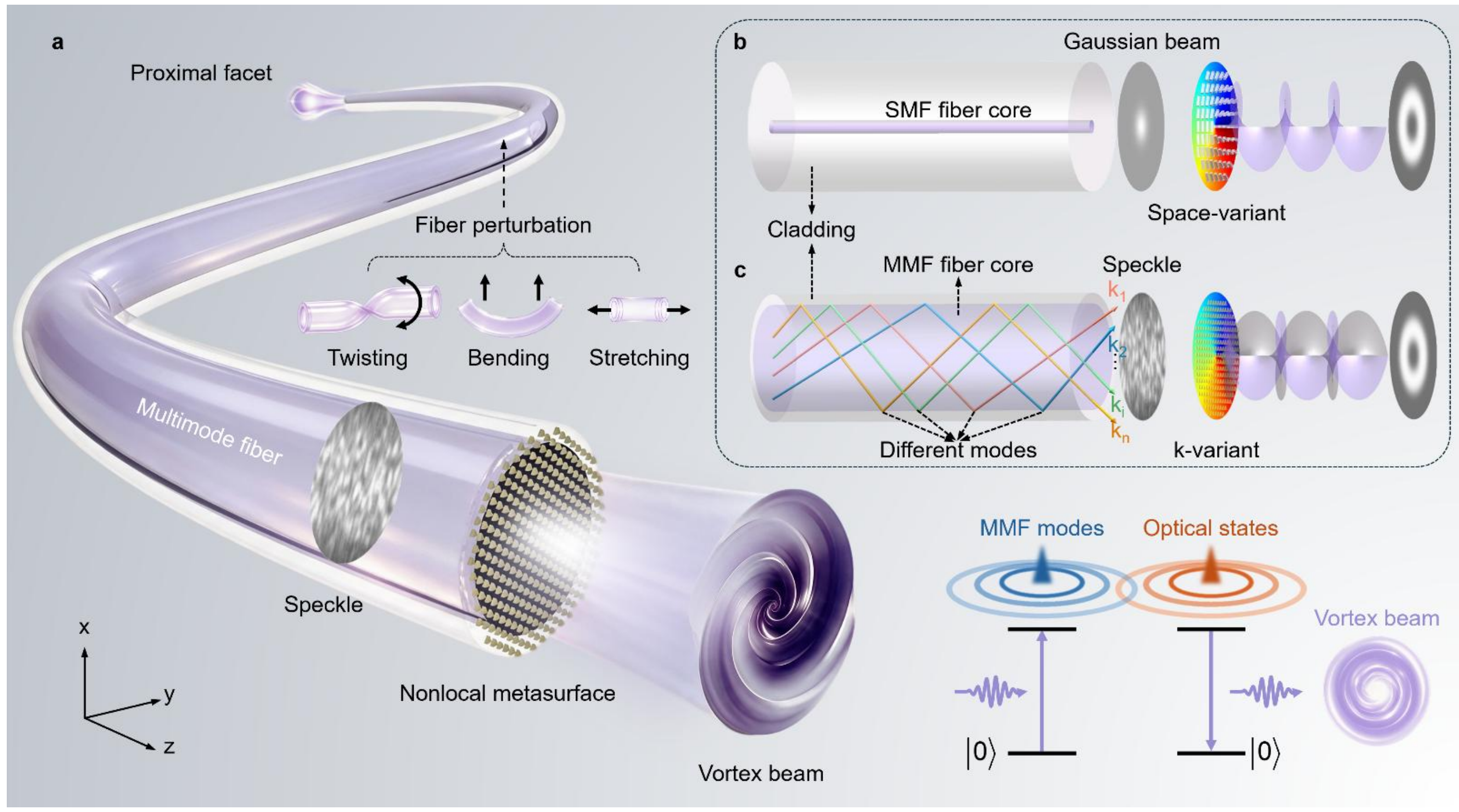

**Fig. 1 Schematic of vortex generation from an MMF enabled by an in-situ integrated high-index dielectric nonlocal Huygens metasurface. a,** The light is incident into the proximal facet of an MMF, and an arbitrary speckle pattern emerges at the fiber end facet. The on-fiber nonlocal metasurface, composed of two-dimensional periodic high-index dielectric nanostructures, functions as a momentum-space wavefront shaping device that transforms the distorted wavefront into vortex beams. The lower-right schematic shows momentum-space coupling between MMF modes and nonlocal modes that possess polarization-vortex nature. **b**,**c**, Conventional SM metafibers (**b**) generate vortex beams via real-space pixelated helical phase modulation, whereas our MMF-integrated nonlocal metasurface (**c**) generates vortex beams via momentum-space coupling with fiber modes of varying wavevectors.

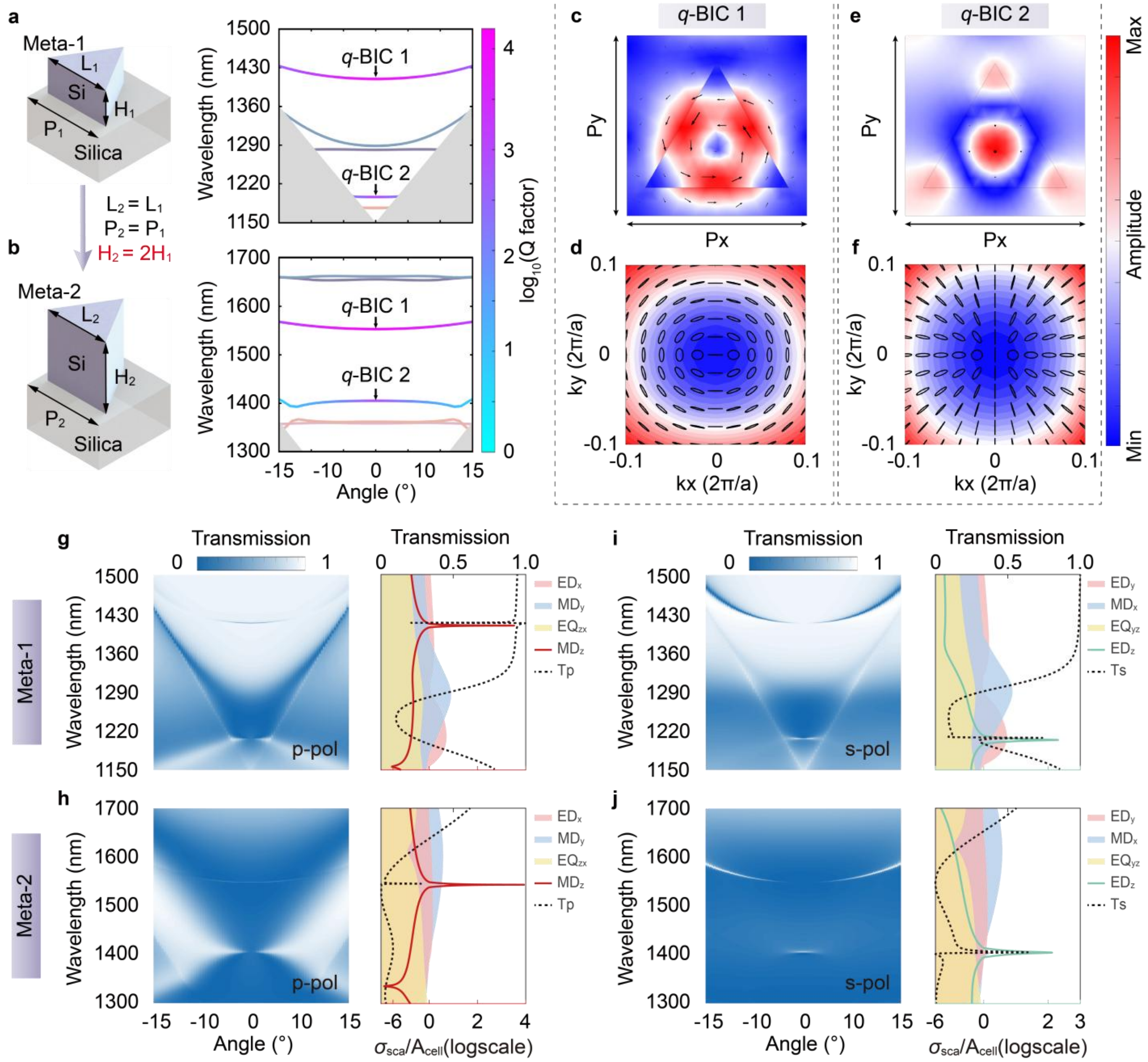

**Fig. 2 Design principles of nonlocal Huygens metasurfaces. a**,**b**, Simulated band structures and Q factors of meta-1 (**a**) and meta-2 (**b**), both consisting of triangular Si nanostructure on a silica substrate. **c-f**, The near-field electric field distribution (**c**, **e**) and far-field momentum-space polarization states (black lines) distribution (**d**, **f**) near the *Γ* point ($k_x = k_y = 0$) for modes *q*-BIC 1 (**c**, **d**) and *q*-BIC 2 (**e**, **f**) of meta-2. **g-j**, Simulated angular spectrum in transmission modes and multipole decompostions of meta-1 (**g**, **i**) and meta-2 (**h**, **j**) under p (**g**, **h**) or s-polarization (**i**, **j**) excitation. ED is the electric dipole, MD is the magnetic dipole, EQ is the electric quadrupole, Tp and Ts are total transmittances under normal incidence with p or s-polarization, respectively.

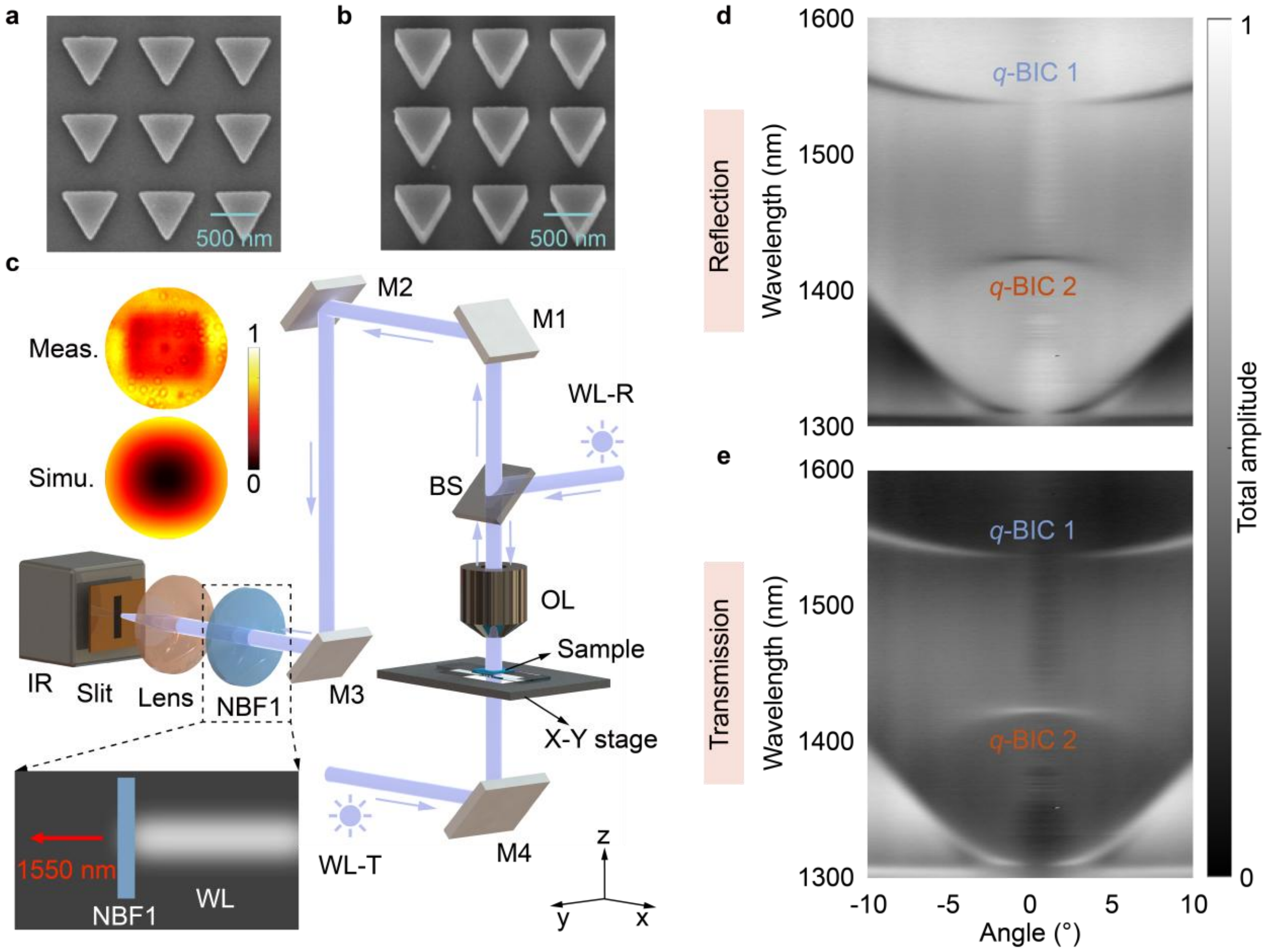

**Fig. 3 High-Q high-transmission bands in a Si nonlocal Huygens metasurface fabricated on a planar quartz substrate. a**,**b**, The top-view (**a**) and side-view (**b**) SEM images of the fabricated sample, scale bar: 500 nm. **c**, The experimental setup (a commercial angle-resolved micro-spectrometer) for characterizing the optical properties of the nonlocal metasurface. M: mirror; OL: objective lens; BS: beam splitter; WL: white light; R: reflection; T: transmission; NBF: narrow band filter. The upper insets present the measured and simulated far-field intensity profiles near *q*-BIC 1 with the emission angle of 12°. The lower inset illustrates the performance of NBF1, transmitting only 1550 nm from a white-light illuminator. **d,e,** The measured angle-resolved reflection (**d**) and transmission (**e**) spectra within the angular range of ±10°, with *q*-BIC 1 and *q*-BIC 2 modes labeled in blue and orange, respectively.

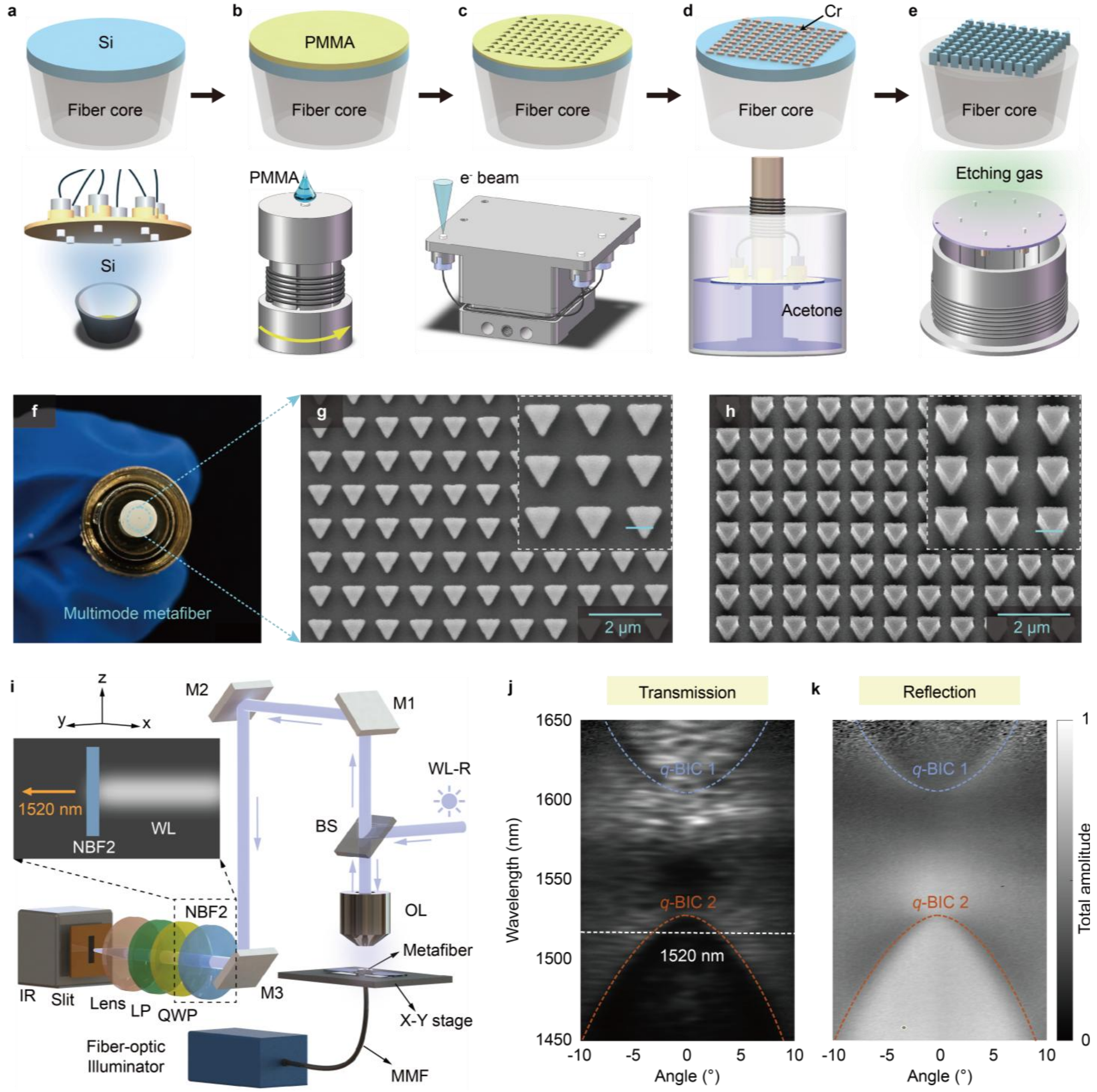

**Fig. 4 Multimode fibers integrated with nonlocal Huygens metasurfaces. a-e**, The first row shows the fabrication process of the fiber-integrated high-index dielectric metasurface, and the second row shows the corresponding CMOS-compatible holders and baseplates used in the fabrication process. **f**, The optical photograph of the MMF-integrated metadevice. **g**,**h**, The top-view (**g**) and side-view (**h**) SEM images of the on-fiber nonlocal metasurface (scale bar: 2 μm, insert scale bar: 500 nm). **i**, The experimental setup for characterizing the optical performance of the MMF-integrated metadevice. M: mirror; OL: objective lens; BS: beam splitter; WL: white light; R: reflection; T: transmission; NBF: narrow band filter. The inset illustrates the performance of NBF2, transmitting only 1520 nm from a white-light illuminator. **j**,**k**, The measured angle-resolved transmission (**j**) and reflection (**k**) spectra of the nonlocal metasurface fabricated onto the MMF end facet, with the blue and orange dashed lines respectively illustrating the *q*-BIC 1 and *q*-BIC 2 bands.

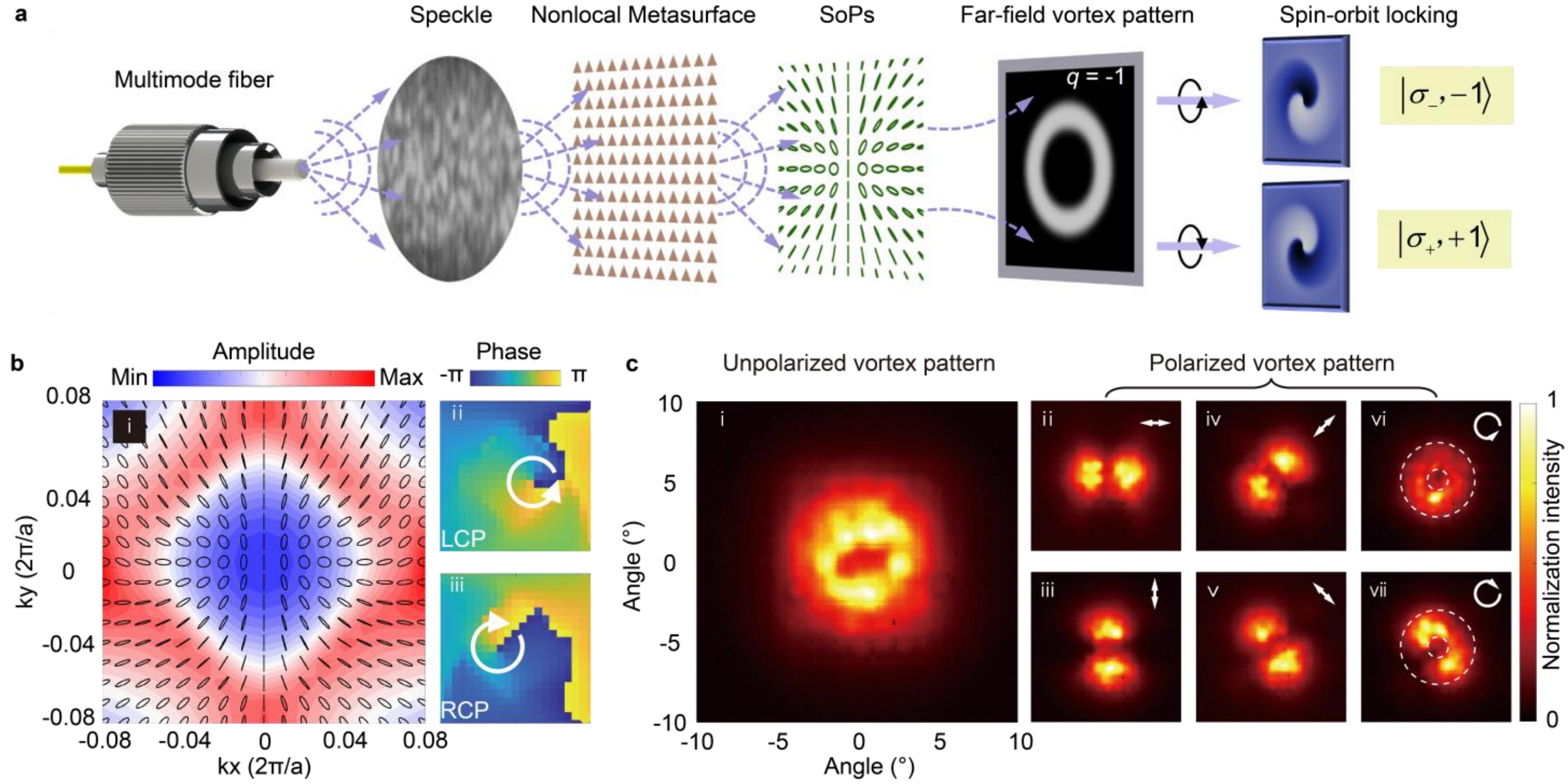

**Fig. 5 Momentum-space vortex beam generation with the MMF-based metadevice. a**, Schematic illustrating the process of generating vortex beams. **b**, The calculated polarization vortex configuration of *q*-BIC 2 (i), and phase distributions for LCP (ii) and RCP (iii) components of *q*-BIC 2. **c**, The measured unpolarized far-field radiation pattern (i), linear polarization (ii-v) and circular polarization (vi and vii) images at the wavelength of 1520 nm.

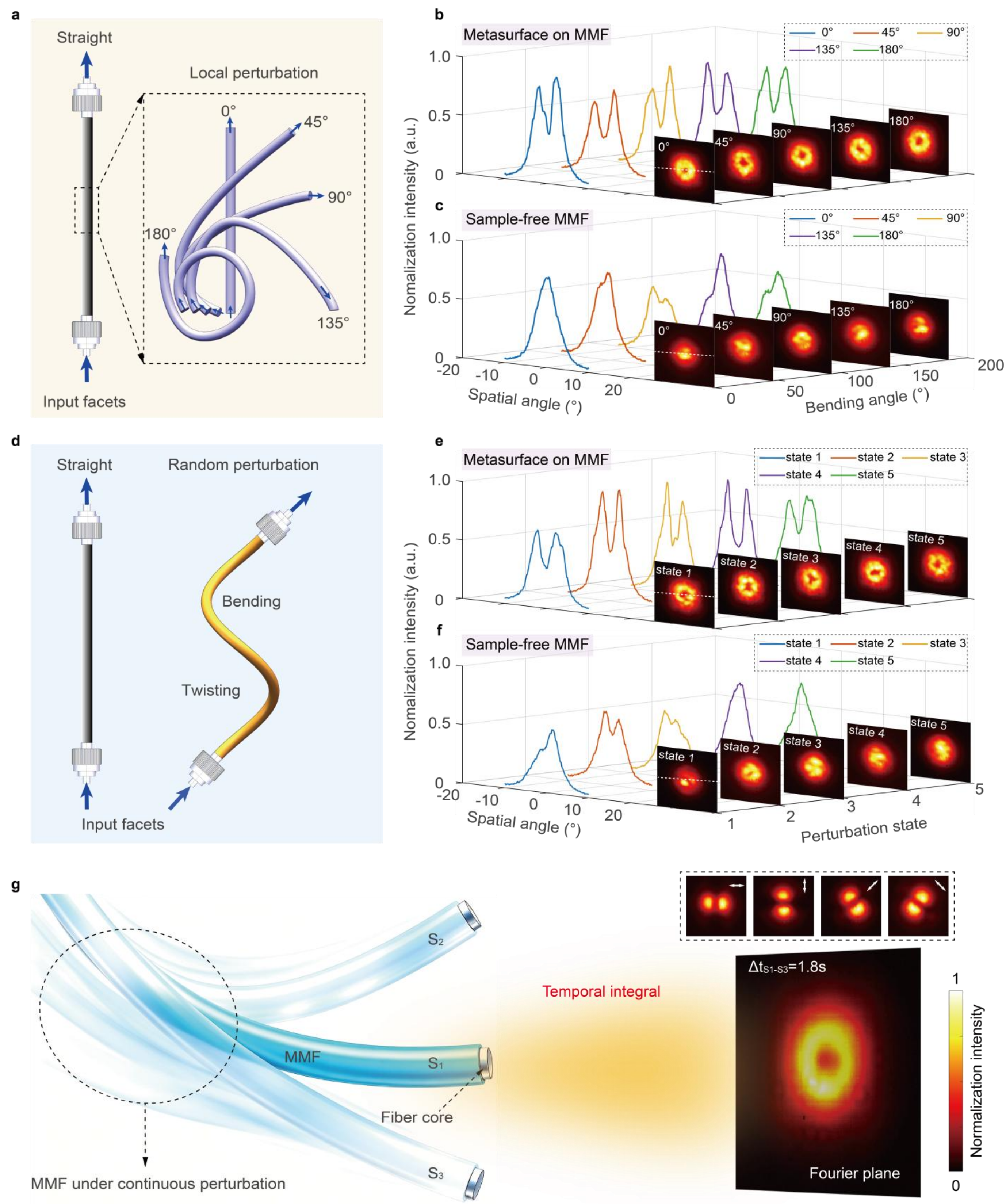

**Fig. 6 Robust optical vortex generation against static and dynamic fiber deformation. a**, Schematic of the bending-angle test used to characterize the device robustness, with the fiber bending angles of 0°, 45°, 90°, 135°, and 180°. **b**,**c**, Corresponding far-field intensity patterns at the Fourier plane and corresponding cross-sectional line profiles (taken along the white dashed lines in the insets) for (**b**) the metasurface-integrated MMF and (**c**) a bare MMF under the different bending states. **d**, Schematic of the static perturbation test (random twisting and bending) used to further characterize the device robustness. **e**,**f**, Corresponding far-field patterns and cross-sectional profiles for (**e**) the metasurface-integrated MMF and (**f**) the bare MMF. **g**, Time-integrated far-field intensity distribution acquired over a 1.8 s exposure period while the MMF-based metadevice underwent continuous dynamic bending and twisting; $S_1$, $S_2$, and $S_3$ denote representative instantaneous snapshots during the stochastic perturbation process.